\documentclass[12pt]{article}
\pdfoutput=1
\usepackage{jheppub}
\usepackage{verbatim}
\usepackage{amsmath}
\usepackage{amssymb}
\usepackage{amsthm}
\usepackage{slashed}
\usepackage{amsfonts}
\usepackage[utf8]{inputenc}

\newcommand{\be}{\begin{equation}}
\newcommand{\ee}{\end{equation}}
\newcommand{\bfig}{\begin{figure}\begin{center}}
\newcommand{\efig}{\end{center}\end{figure}}
\newcommand{\bi}{\begin{itemize}}
\newcommand{\ei}{\end{itemize}}

\newcommand{\lan}{\langle}
\newcommand{\ran}{\rangle}
\newcommand{\Tr}{\mathrm{Tr}}

\theoremstyle{definition}

\begin{document}
\title{The Factorization Problem in Jackiw-Teitelboim Gravity}
\author[a]{Daniel Harlow}
\author[b]{Daniel Jafferis}
\affiliation[a]{Center for Theoretical Physics, Massachusetts Institute of Technology, Cambridge MA, 02138 USA}
\affiliation[b]{Center for the Fundamental Laws of Nature, Harvard University, Cambridge MA 02138 USA}
\emailAdd{harlow@mit.edu}
\emailAdd{jafferis@physics.harvard.edu}
\abstract{In this note we study the $1+1$ dimensional Jackiw-Teitelboim gravity in Lorentzian signature, explicitly constructing the gauge-invariant classical phase space and the quantum Hilbert space and Hamiltonian.  We also semiclassically compute the Hartle-Hawking wave function in two different bases of this Hilbert space.  We then use these results to illustrate the gravitational version of the factorization problem of AdS/CFT: the Hilbert space of the two-boundary system tensor-factorizes on the CFT side, which appears to be in tension with the existence of gauge constraints in the bulk.  In this model the tension is acute: we argue that JT gravity is a sensible quantum theory, based on a well-defined Lorentzian bulk path integral, which has no CFT dual.  In bulk language, it has wormholes but it does not have black hole microstates.  It does however give some hint as to what could be added to to rectify these issues, and we give an example of how this works using the SYK model.  Finally we suggest that similar comments should apply to pure Einstein gravity in $2+1$ dimensions, which we'd then conclude also cannot have a CFT dual, consistent with the results of Maloney and Witten.}
\maketitle
\section{Introduction}
Most discussions of bulk physics in AdS/CFT focus on perturbative fields about a fixed background \cite{Witten:1998qj,Gubser:1998bc,Banks:1998dd,Aharony:1999ti,Hamilton:2006az,Heemskerk:2009pn,Heemskerk:2012mn,Heemskerk:2012np}.  This has led to much progress in understanding the correspondence, see \cite{Harlow:2018fse} for a recent review, but sooner or later we will need to confront the fact that the bulk theory is gravitational; in generic states gravitational backreaction cannot be treated as an afterthought.  In particular the strong redshift effects near black hole horizons make physics from the point of view of the outside observer unusually sensitive to gravitational effects there \cite{Dray:1984ha,Shenker:2013pqa,Shenker:2013yza,Shenker:2014cwa,Polchinski:2015cea}.  Gravitational backreaction also provides the mechanism by which the holographic encoding of the higher-dimensional bulk into the lower-dimensional boundary theory breaks down if we try to preserve bulk locality beyond what is allowed by holographic entropy bounds \cite{Almheiri:2014lwa,Pastawski:2015qua,Harlow:2016vwg,Harlow:2018fse}. 


\bfig
\includegraphics[height=4cm]{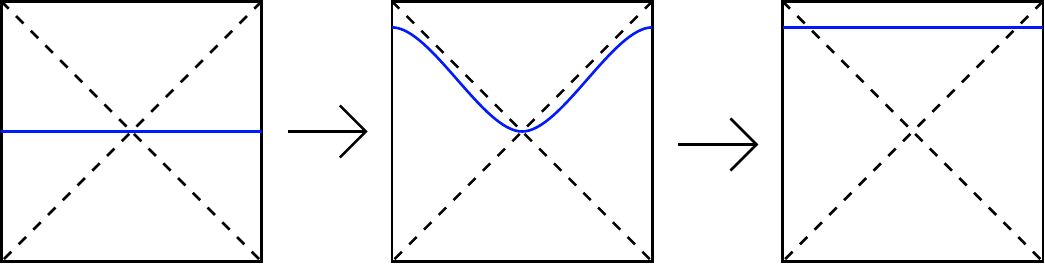}
\caption{Different kinds of time evolution in the bulk and boundary: the endpoints of a bulk timeslice are moved up and down using the ADM Hamiltonian, while the interior of the slice is evolved using the Hamiltonian constraint.  Since the Hamiltonian constraint is zero on physical states, the second two slices describe the same CFT state.}\label{timeslicesfig}
\efig
One especially confusing aspect of gravitational physics is that time translations are gauge transformations: much of the interesting dynamics is tied up in the gauge constraints.  For example consider figure \ref{timeslicesfig}.  It is sometimes said that in the context of the two-sided AdS-Schwarzschild geometry, we can see the interior of the wormhole by evolving both boundary times forward \cite{Hartman:2013qma,Susskind:2014rva,Susskind:2015toa,Brown:2015bva,Brown:2015lvg}.  But in fact as we move from the left to the central diagram in figure \ref{timeslicesfig}, we see that we can evolve the boundary times as far to the future as we like without ever having the bulk time slice go behind the horizon.  It is not until we move the interior part of the slice up that we start to directly see physics behind the horizon, but this is precisely the part of the evolution which is generated by the Hamiltonian constraint of general relativity.  How are we to distinguish the slices in the center and right diagram, when from the CFT point of view they describe precisely the same quantum state?  

A related point is that the entire formation and evaporation of a small black hole in AdS is spacelike to some boundary time slice, and thus must be describable purely via the Hamiltonian constraints. In other words, there is a spatial slice that intersects the collapsing matter prior to the formation of an event horizon and another spatial slice that intersects the Hawking radiation after the complete evaporation of the resulting black hole, both of which asymptote to the same time slice of the boundary. Such a description of the Hawking process would be complementary to the more standard one in which temporal diffeomorphisms are imagined to be gauge fixed, directly tying together the bulk and boundary time evolutions. 

Another interesting question related to the black hole interior and gauge constraints is the following. Say that we believe that a black hole which evolves for a long enough time develops a firewall \cite{Braunstein:2009my,Almheiri:2012rt}. Where precisely does it form?  A naive answer would be at the event horizon, but this is unlikely to actually be correct. The event horizon is a teleological notion, which for example can be modified by putting our evaporating black hole inside of a huge shell of collapsing matter, which will not collapse until long after our black hole evaporates.  It seems doubtful, to say the least, that we could remove a firewall by so silly a trick as this.  One might also suggest that firewalls form at ``the'' apparent horizon, but actually apparent horizons are highly non-unique since they depend on a choice of Cauchy slice \cite{Eardley:1997hk}. The recently studied ``holographic screens'' \cite{Bousso:2015mqa} also are too non-unique to do the job. If indeed there are firewalls, there should be a gauge-invariant prescription for where (and also how) they form.\footnote{There is clearly at least some approximate sense of ``where'' the edge of a black hole currently is, for example the event horizon telescope will soon image the disc of Sagittarius A* and the LIGO team already simulates black hole merger events using code which excises some kind of black hole region. It would be interesting to understand the generality of the underlying assumptions in such calculations, and whether or not a formal definition could be given which applies in sufficiently generic situations to be relevant for the firewall arguments.}   

In this paper we will be primarily interested in a third issue raised by considering gravitational physics in AdS/CFT: the \textit{factorization problem} \cite{Harlow:2015lma,Guica:2015zpf}.  This is the observation that the presence of gauge constraints in the bulk poses a potential obstacle to the existence of a CFT dual, since such constraints might not be consistent with the tensor product structure of the boundary field theory when studied on a disconnected spacetime.  As a simple example, consider $1+1$ dimensional Maxwell theory on a line interval times time:
\be
S=-\frac{1}{4}\int_{-\infty}^\infty dt\int_0^L dx F_{\mu\nu}F^{\mu\nu}.
\ee  
The equations of motion for this theory tell us that the electric field is constant throughout spacetime, but its value cannot be the only dynamical variable since phase space is always even-dimensional.  To find the other dynamical variable we need to be more careful about the boundary conditions: the variation of the Maxwell action on any spacetime $M$ has a boundary term
\be
\delta S\supset -\int_{\partial M}\sqrt{\gamma}r_\mu F^{\mu\nu}\delta A_\nu,
\ee
where $r_\mu$ is the (outward pointing) normal form. To formulate a good variational problem, we need to impose boundary conditions such that this term vanishes for variations within the space of configurations obeying the boundary conditions.  There are various options for these boundary conditions, the natural choice for AdS/CFT (the ``standard quantization'') is to take
\be
A_\mu|_{\partial M} \propto r_\mu.
\ee
These boundary conditions are not preserved under general gauge transformations: we must at least require that any gauge transformation $\Lambda(x)$ approaches a constant on each connected component of $\partial M$.  In fact the most natural choice is to require that these constants are all zero for the gauge transformations which we actually quotient by: transformations where they are nonzero are then viewed as asymptotic symmetries which act nontrivially on phase space.\footnote{In AdS/CFT this is motivated by wanting to preserve boundary locality.  Quotienting by gauge transformations which approach nonzero constants amounts to imposing a singlet condition for a global symmetry in the boundary CFT, which violates locality. When the boundary is $0+1$ dimensional locality is a less serious constraint, but we still want to avoid projections that mix the two asymptotic boundaries.  Quotienting by gauge transformations which approach nonzero constants then requires us to just set $E=0$, which leads to an empty theory.}  For our example on $\mathbb{R}\times I$, it is always possible to go to $A_0=0$ gauge by a gauge transformation that vanishes at the endpoints of the interval.  The equation of motion then requires that
\be
A_x=-E t+a,
\ee
where $a$ is a constant which could be removed by an ``illegal'' gauge transformation $\Lambda=-ax$.  Since we are not allowed to do such gauge transformations, $a$ is physical: in fact it is nothing but the Wilson line from $x=0$ to $x=L$ at $t=0$.  After quantization, this system just becomes the quantum mechanics of a particle on a circle (here we are assuming the gauge group is $U(1)$, not $\mathbb{R}$), and in particular it has no tensor product decomposition into degrees of freedom to the left and right of the line $x=L/2$.

Of course pure Maxwell theory is not expected to have a gravity dual anyways, so the non-factorization of this system may at first appear uninteresting.  But in fact it has far-reaching consequences: the Einstein-Maxwell theory on the two-sided AdS-Schwarzschild geometry in any spacetime dimension has a zero mode sector which is equivalent to this theory, and which tells us, among other things, that in gravitational theories with CFT duals, one-sided states must exist with all gauge charges allowed by charge quantization \cite{Harlow:2015lma,hotop}.  Nonetheless it would be nice to concretely realize the factorization problem in a gravitational model, which at least somewhat plausibly might have been hoped to have a CFT dual.  Our main goal in this paper is to do precisely this.  

The theory we will study is the $1+1$ dimensional Jackiw-Teitelboim theory of dilaton gravity \cite{Jackiw:1984je,Teitelboim:1983ux,Almheiri:2014cka}, with bulk Lagrangian density
\be
\mathcal{L}=\Phi_0R+\Phi(R+2).
\ee
The first two sections of our paper will simply repeat the analysis sketched above for Maxwell theory in this model, which we will see similarly does not have a factorized Hilbert space.\footnote{A similar analysis of the asymptotically-Minkowski CGHS model was done in \cite{Kuchar:1996zm}. See also \cite{Nickel:2010pr,Crossley:2015tka}, who discussed degrees of freedom related to those we find here in a one-sided context.}  This lack of factorization implies that the theory cannot have a CFT dual, nevertheless it is a self-consistent quantum mechanical system, albeit one with a continuous spectrum. There is nothing in the gravitational analysis that requires a breakdown of the JT description. However, we will comment on what might be added to the theory so that it could have a CFT dual; then the JT Lagrangian would be a low energy approximation and the canonical gravity analysis would eventually exit its regime of validity.

There has been considerable recent interest in this model, see \cite{Sarosi:2017ykf} for a nice review and further references.  Our approach however is rather different in method and emphasis from this literature:
\bi
\item We work primarily in Lorentzian signature, focusing on identifying the physical on-shell degrees of freedom. 
\item The ``Schwarzian'' Lagrangian will make no appearance in our analysis.  Indeed the Schwarzian theory is not sensible by itself in Lorentzian signature, from our point of view it is an artifact of a particular way of evaluating the Euclidean path integral.  
\item We will make almost no mention of the group $SL(2,\mathbb{R})$, which acts on the JT theory neither as a global symmetry nor as a natural subgroup of the gauged diffeomorphism group.  
\ei
These differences arise because our analysis is not particularly motivated by the SYK model \cite{Sachdev:1992fk,Kitaev,Polchinski:2016xgd,Maldacena:2016hyu,Kitaev:2017awl}, while the Schwarzian action on elements of $\mathrm{diff}(S^1)/SL(2,\mathbb{R})$ is particularly well-suited for understanding how the JT theory is embedded in that model \cite{Kitaev,Maldacena:2016hyu,Jensen:2016pah,Engelsoy:2016xyb,Maldacena:2016upp,Kitaev:2017awl}.  We nonetheless  include a section where we explain how our analysis fits into the Lorentzian version of the SYK model, and we will there explain how to understand our results in the Schwarzian language.  We hope that our analysis of the JT theory with ``more conventional'' techniques will be useful even to SYK-oriented readers.

Finally we discuss some of the possible implications of our work for higher-dimensional gravity.  In particular, we will argue that there is a quite close analogy between JT gravity in $1+1$ dimensions and pure Einstein gravity in $2+1$ dimensions: both seem to have precise path integral descriptions in the bulk, both have wormhole solutions, both have a two-sided Hilbert space which does not factorize, neither have black hole microstates counted by the Bekenstein-Hawking formula, and neither have CFT duals.  In both cases the answers to these questions become more standard once matter is added, something we leave for future work.  

Previous attempts to quantize Jackiw-Teitelboim gravity with other boundary conditions include \cite{Henneaux:1985nw,NavarroSalas:1992vy,Constantinidis:2008ty}.

\section{Classical Jackiw-Teitelboim gravity}
The Jackiw-Teitelboim action on a $1+1$ dimensional asymptotically-AdS spacetime $M$ is given by
\be\label{JTact}
S=\Phi_0\left(\int_M d^2 x \sqrt{-g}R+2\int_{\partial M}\sqrt{|\gamma|}K\right)+\int_M d^2 x \sqrt{-g}\,\Phi\left(R+2\right)+2\int_{\partial M} dt \sqrt{|\gamma|} \Phi (K-1).
\ee
Here $\Phi_0$ is a large positive constant, which in a situation where we obtained this theory by dimensional reduction would correspond to the volume in higher-dimensional Planck units of the compact directions \cite{Almheiri:2014cka,Nayak:2018qej}.  From a two-dimensional point of view $\Phi_0$ is just the coefficient of the topological Einstein-Hilbert part of the action.  $\Phi$ is a dynamical scalar field we will call the dilaton.  $K$ is the trace of the extrinsic curvature of the boundary, defined as
\be
K\equiv \gamma^{\mu\nu} \nabla_\mu r_\nu,
\ee 
with $\gamma_{\mu\nu}$ the induced metric on the boundary and $r_\mu$ the outward-pointing normal form there.\footnote{In our conventions the normal vector $r^\mu$ is outward-pointing if it is spacelike but inward-pointing if it is timelike.  This ensures that Stokes theorem
\be
\int_M d^d x \sqrt{|g|} \nabla_\mu V^\mu=\int_{\partial M} d^{d-1}x \sqrt{|\gamma|}r_\mu V^\mu
\ee
holds regardless of the signature of the boundary.  The induced metric is related to the ordinary one by $\gamma_{\mu\nu}\equiv g_{\mu\nu}\mp r_\mu r_\nu$, where $r_\mu$ is spacelike/timelike.}  The boundary term not involving $K$ is a holographic renormalization, which ensures that the action and Hamiltonian are finite on configurations obeying the boundary conditions we will soon discuss.  
The variation of this action is\footnote{For spacetimes with additional boundaries which are not asymptotically-AdS, such as the time slice $\Sigma$ we will use in computing the Hartle-Hawking wave function below, this equation remains correct except that the terms $-\int_{\partial M} dx\sqrt{|\gamma|}\left(2\delta\Phi+\Phi\gamma^{\alpha\beta}\delta\gamma_{\alpha\beta}\right)$ appear only on the asymptotically-AdS parts of the boundary, since it is only there that we include the holographic renormalization counterterm $-2\int dx \sqrt{|\gamma|}\Phi$.}
\begin{align}\nonumber
\delta S=&\int d^2 x \sqrt{-g}\Big[\left(\frac{1}{2}(R+2)\Phi g^{\mu\nu}-R^{\mu\nu}\Phi+\nabla^\mu\nabla^\nu\Phi-g^{\mu\nu}\nabla^2 \Phi\right)\delta g_{\mu\nu}\\\nonumber
&-\Phi_0\left(R^{\mu\nu}-\frac{1}{2}R g^{\mu\nu}\right)\delta g_{\mu\nu}+(R+2)\delta \Phi \Big]\\
&+\int_{\partial M} dx \sqrt{|\gamma|} \Big[2 (K-1) \delta \Phi+\left(r^\nu \nabla_\nu \Phi-\Phi\right)\gamma^{\alpha \beta}\delta \gamma_{\alpha \beta}\Big],\label{var}
\end{align}
so after some simplification the equations of motion are
\begin{align}\nonumber
R+2&=0\\\label{eom}
\left(\nabla_\mu \nabla_\nu -g_{\mu\nu}\right)\Phi&=0.
\end{align}
As in the electromagnetic case, we need to choose boundary conditions such that the boundary terms in \eqref{var} vanish for any variation in the space of configurations obeying these boundary conditions.  The obvious choice is to fix the induced metric $\gamma_{\mu\nu}$ and dilaton $\Phi$ at the AdS boundary, which we can do by imposing 
\begin{align}\nonumber
\gamma_{tt}|_{\partial M}&=r_c^2\\
\Phi|_{\partial M}&=\phi_b r_c \label{bc}
\end{align}
and then taking $r_c\to\infty$ with $\phi_b$ fixed and positive.  $\phi_b$ is analogous to the AdS radius in Planck units in higher dimensions, it will be large in the semiclassical limit.  These boundary conditions are only preserved by the subset of infinitesimal diffeomorphisms $\xi^\mu$ which approach an isometry of the boundary metric,
\be
\gamma_\mu^{\phantom{\mu}\alpha}\gamma_\nu^{\phantom{\nu}\beta}\nabla_{(\alpha}\xi_{\beta)}|_{\partial M}=0,
\ee
which means that the pullback of $\xi_\mu$ to each component of $\partial M$ must be a time translation.  As in electromagnetism, we will only actually quotient by diffeomorphisms where these time translations are trivial, with the motivation again being to preserve boundary locality (note also that otherwise we would be left with a boundary theory with no states of nonzero energy).  We thus expect boundary time translations to be asymptotic symmetries which act nontrivially on phase space: indeed they will be generated by the ADM Hamiltonians on the respective boundaries.  

To understand these Hamiltonians more concretely, we can then define a ``CFT metric'' at each boundary,  
\be
\gamma_{\mu\nu}^{CFT}\equiv \frac{\gamma_{\mu\nu}}{r_c^2}, 
\ee
in terms of which we can define a boundary stress tensor \cite{Balasubramanian:1999re}
\be
T_{CFT}^{\mu\nu}\equiv \frac{2}{\sqrt{|\gamma^{CFT}|}}\frac{\delta S}{\delta \gamma_{\mu\nu}^{CFT}}.
\ee
From \eqref{var} we then apparently have
\be\label{TCFT}
T_{CFT}^{\mu\nu}=2r_c^3\gamma^{\mu\nu}\left(r^\lambda \nabla_\lambda \Phi-\Phi\right)|_{\partial M}.
\ee
The $tt$ component of this (which is the only component) at each boundary is the AdS analogue of the ADM Hamiltonian for that boundary, and the full canonical Hamiltonian is the sum of these Hamiltonians. From now on we specialize to boundary conditions where there are precisely two asymptotically-AdS boundaries: we are then restricting to spacetimes with topology $\mathbb{R}\times [0,1]$, on which the metric and dilaton obey \eqref{bc} at each asymptotic boundary, and the full Hamiltonian $H$ is the sum of left and right ADM Hamiltonians $H_L$ and $H_R$.

\subsection{Solutions}
There are various ways to describe the set of solutions of the equations of motion \eqref{eom}.  One nice way is to observe that the first equation requires the metric to have constant negative curvature, which means that it is described by a piece of $AdS_2$.  $AdS_2$ can be obtained via an embedding into $1+2$ dimensional Minkowski space, with metric
\be\label{embedmet}
ds^2=-dT_1^2-dT_2^2+dX^2.
\ee
$AdS_2$ is the universal cover of the induced geometry on the surface
\be\label{embedding}
T_1^2+T_2^2-X^2=1
\ee
in this Minkowski space.  The two $AdS$ boundaries are at $X\to \pm \infty$.

We may then ask what the set of possible solutions for $\Phi$ look like: the answer is that for any solution of \eqref{eom}, the slices of constant $\Phi$ are given by the intersections of the embedding surface \eqref{embedding} with a family of hyperplanes
\be
\Phi=AT_1+BT_2+CX,
\ee
where $A, B, C$ are three fixed real parameters which label the solution: we can think of them as parametrizing the normal vector $n^{\mu}=(-A,-B,C)$ to the hypersurfaces.  The solutions where $n^\mu$ is spacelike or null will never obey our boundary conditions \eqref{bc}, since $\Phi$ will be negative almost everywhere on one of the AdS boundaries at $X\to\pm \infty$.  When $n^\mu$ is timelike, we can set $B=C=0$ by an $SO(1,2)$ rotation in the embedding space, so we can restrict to solutions of the form
\be
\Phi=\Phi_h T_1,
\ee
where we have relabelled $A$ to $\Phi_h$ for a reason which will be apparent momentarily.

\bfig
\includegraphics[height=7cm]{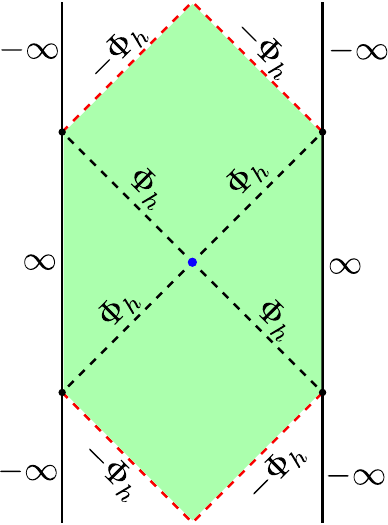}
\caption{The dilaton profile in the wormhole solution of Jackiw-Teitelboim gravity.  Between the two vertical black lines the geometry is global $AdS_2$, and we indicate the value of the dilaton on various surfaces.  In Reissner-Nordstrom language, the dashed black lines where $\Phi=\Phi_h$ are the ``outer horizon'', while the dashed red lines where $\Phi=-\Phi_h$ are the ``inner horizon''.  The dynamical problem with boundary conditions \eqref{bc} is well-defined only in the shaded green region.  If we assume that the inner horizon is singular, this solution describes a wormhole connecting two asymptotically-$AdS$ boundaries.}\label{solfig}
\efig
We can present these solutions more concretely by choosing coordinates
\begin{align}\nonumber
T_1&=\sqrt{1+x^2}\cos \tau\\\nonumber
T_2&=\sqrt{1+x^2}\sin \tau\\
X&=x,
\end{align}
in terms of which we have
\begin{align}\nonumber
ds^2&=-(1+x^2)d\tau^2+\frac{dx^2}{1+x^2}\\
\Phi&=\Phi_h\sqrt{1+x^2}\cos \tau.\label{globalsol}
\end{align}
We illustrate this solution in figure \ref{solfig}.  Its maximal extension involves infinitely many boundary regions, some with $\Phi=+\infty$ and some with $\Phi=-\infty$. As is normal with Reissner-Nordstrom-type solutions, we expect that small matter fluctuations (once matter is included) will cause the ``inner horizons'', where $\Phi=-\Phi_h$, to become singular, collapsing the geometry down to just the wormhole region shaded green in figure \ref{globalsol}. In pure Jackiw-Teitelboim gravity there is no matter which can do this, but the dynamical problem with two asymptotically-AdS boundaries obeying \eqref{bc} is still only well-defined in the green region, since additional boundary data would be needed to extend the solution out of this region.  Since we are primarily interested in constructing a theory which is a good model for gravity in higher dimensions, where the inner horizon is indeed always singular, we find it simplest to just truncate the spacetime at the inner horizon.\footnote{Were we not to do this, then we would need to include additional degrees of freedom to keep track of what happens on the other pieces of the boundary.}

In addition to these ``global'' coordinates, we can also go to ``Schwarzschild'' coordinates, via
\begin{align}\nonumber
T_1&=r/r_s\\\nonumber
T_2&=\sqrt{(r/r_s)^2-1}\sinh (r_s t)\\
X&=\sqrt{(r/r_s)^2-1}\cosh (r_s t),\label{schwarz}
\end{align}
in terms of which we have
\begin{align}\nonumber
ds^2&=-(r^2-r_s^2)dt^2+\frac{dr^2}{r^2-r_s^2}\\
\Phi&=\phi_b r.\label{schsol}
\end{align}
For $r>r_s$ and $-\infty<t<\infty$, these coordinates cover the ``right exterior'' piece of the green shaded region in figure \ref{solfig}, which as usual lies between the right asymptotic boundary and the right part of the $\Phi=\Phi_h$ bifurcate outer horizon.  The parameters of these solutions are related via
\be
r_s=\frac{\Phi_h}{\phi_b}.
\ee
These coordinates have the nice feature that slices of constant $r$ are also slices of constant $\Phi$, so in particular the cutoff surface in the boundary conditions \eqref{bc} just lies at $r=r_c$, and moreover $t$ becomes the boundary time.  

From \eqref{schsol} it is easy to evaluate the boundary stress tensor \eqref{TCFT} on each boundary for these solutions, one finds
\be\label{Hs}
H_L=H_R=\frac{\Phi_h^2}{\phi_b},
\ee
so the full canonical Hamiltonian evaluates to
\be\label{Htot}
H=H_L+H_R=\frac{2\Phi_h^2}{\phi_b}.
\ee

\subsection{Phase space and symplectic form}
\bfig
\includegraphics[height=4cm]{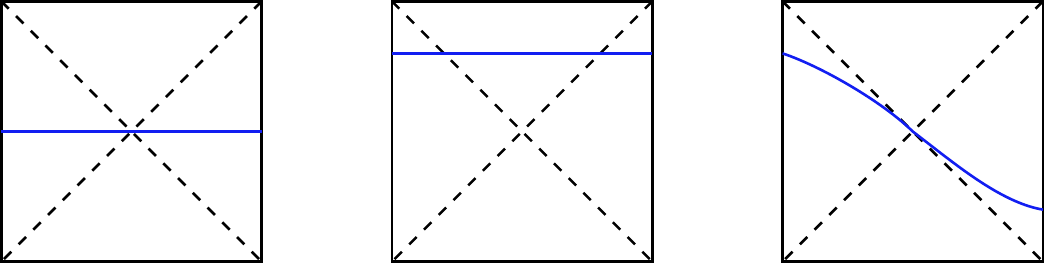}
\caption{Different time slices of the wormhole solution can correspond to different initial data for the JT gravity.  Here the first and second slices are different points in phase space, while the first and third are the same since they differ by evolution by $H_R-H_L$.}\label{timeslices2fig}
\efig
In the previous subsection we described a one-parameter family of solutions of the JT theory, labeled by the value of the dilaton on the bifurcate horizon, $\Phi_h$.  This parameter is analogous to the electric field in our $1+1$ Maxwell example: it is locally measurable.  As in the Maxwell example however, $\Phi_h$ cannot be the only parameter on the space of solutions: phase space must be even-dimensional.  The other parameter, analogous to $a$ in the Maxwell example, arises because in going to the coordinates \eqref{globalsol}, we in fact did an illegal gauge transformation.  The easiest way to restore any solution parameters we removed this way is to act with another illegal gauge transformation, of the class which approaches an asymptotic symmetry at infinity: the parameters of this gauge transformation (modulo legal gauge transformations) will become the gravitational analogue of $a$ in the electromagnetic example.  In the present discussion, the only asymptotic symmetries are time translations on the left and right asymptotic boundaries.  So at first it might seem that we have discovered two new parameters: our phase space still seems odd-dimensional!  But in fact equation \eqref{Hs} tell us that $H_L=H_R$ on all solutions, so the operator $H_L-H_R$ generates no evolution on phase space.  Thus we have only one new parameter, which we will call $\delta$, which tells us how long we evolved the solution \eqref{globalsol} by $H_R+H_L$. More explicitly, the relationship between global time $\tau$ and the ``left'' and ``right'' Schwarzschild times $t_L$, $t_R$ at the AdS boundaries is
\be\label{taut}
\cos \tau =\frac{1}{\cosh (r_st_L)}=\frac{1}{\cosh (r_s t_R)},
\ee
so a slice which is attached to the left and right boundaries at $t_L$ and $t_R$ respectively has
\be\label{delt}
\delta=\frac{t_L+t_R}{2}.
\ee
We illustrate this in figure \ref{timeslices2fig}.

\bfig
\includegraphics[height=4cm]{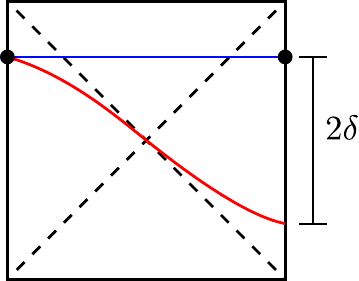}
\caption{Using a geodesic, shown in red, to measure $\delta$.}\label{timeslices3fig}
\efig
There is another somewhat more operational way of describing $\delta$, shown in figure \ref{timeslices3fig}.  The idea is to start at the point on the left boundary where our time slice is attached, fire a geodesic into the bulk which is orthogonal to surfaces of constant $\Phi$, and then see at what time $\hat{t}_R$ this geodesic arrives at the right boundary.  We then have
\be
t_R-\hat{t}_R=2\delta,
\ee   
where $t_R$ is the time where our time slice intersects the right boundary.  Thus we can think of $\delta$ as measuring the ``relative time shift'' between the two boundaries: from now on we will refer to it as the ``time shift operator''  From this point view, the time shift operator is quite similar to the one-sided ``hydrodynamic modes'' discussed in \cite{Nickel:2010pr,Crossley:2015tka}. 

We thus have arrived at the following two-dimensional Hamiltonian system:
\begin{align}\nonumber
\dot{\delta}&=1\\\nonumber
\dot{\Phi}_h&=0\\
H&=\frac{2\Phi_h^2}{\phi_b}.\label{hamsys}
\end{align}
The ranges of these phase space coordinates are $\Phi_h>0$, $-\infty<\delta<\infty$.  For any Hamiltonian system the symplectic form $\omega_{ab}$ is defined by
\be
\dot{x}^a=(\omega^{-1})^{ba}\partial_b H,
\ee
so from \eqref{hamsys} we apparently have
\be
\omega=\frac{4\Phi_h}{\phi_b}d\delta\wedge d \Phi_h,
\ee
which is more elegantly written by changing coordinates from $\Phi_h$ to $H$, giving us
\be\label{Hsymp}
\omega=d\delta \wedge dH.
\ee
Thus $\delta$ is simply the canonical conjugate of $H$.  

Before moving on to the quantum theory, it is convenient to here introduce another pair of coordinates on this phase space.  Roughly speaking these are the geodesic distance between the two endpoints of a time slice and its canonical conjugate, but since that distance is infinite in the $r_c\to\infty$ limit we need to be a bit more careful.  We will defined a ``renormalized geodesic distance'', $L$, via
\be
L\equiv L_{bare}-2\log(2 \Phi|_{\partial M}).
\ee
Using the symmetry generated by $H_R-H_L$ we can always choose $t_L=t_R$, so then from \eqref{globalsol}, we have
\be
L_{bare}=\int_{-x_c}^{x_c} \frac{dx}{\sqrt{1+x^2}},
\ee
with $x_c$ determined in terms of $r_c$ by solving
\be
\phi_b r_c=\Phi_h \sqrt{1+x_c^2}\cos \tau
\ee
and $\tau$ determined in terms of $\delta$ via \eqref{taut} and \eqref{delt}.  We then find
\be\label{Leq}
L=2\log \left(\frac{\cosh \left(\Phi_h\delta/\phi_b\right)}{\Phi_h}\right)=2\log \left(\cosh \left(\sqrt{\frac{E}{2\phi_b}}\delta\right)\right)-\log \frac{\phi_b E}{2},
\ee
which shows that $L$ is indeed a well-defined function on our two-dimensional phase space.  A calculation then shows that if we define
\be
P=2\Phi_h\tanh\left(\Phi_h\delta/\phi_b\right)=\sqrt{2\phi_bE}\tanh\left(\sqrt{\frac{E}{2\phi_b}}\delta\right),
\ee
the symplectic form becomes
\be
\omega=dL\wedge dP,
\ee
so $L$ and $P$ are canonically conjugate variables, both ranging from $-\infty$ to $\infty$.  The Hamiltonian takes a very nice form in terms of $L$ and $P$:\footnote{We thank Henry Lin for pointing out an error in the first version of this paper, which originated in equation \eqref{schwarz} and led to a much more unpleasant formula for the Hamiltonian.}
\be\label{LPH}
H=\frac{P^2}{2\phi_b}+\frac{2}{\phi_b}e^{-L}.
\ee 
In terms of the renormalized geodesic length, JT gravity becomes just the mechanics of a non-relativistic particle moving in an exponential potential!  This is a scattering problem, with waves that come in from $L=\infty$ and reflect off of the potential, and indeed that is what happens in the solutions \eqref{globalsol}.  

\section{Quantum Jackiw-Teitelboim gravity}\label{quantsec}
We now discuss the quantization of JT Gravity, starting with the Hilbert space formalism.
\subsection{Hilbert Space and energy eigenstates}\label{hilbsec}
The most straightforward proposal for the Hilbert space of the quantum JT theory is that it is spanned by a set of delta-function normalized states $|E\ran$, with $E>0$, such that
\be
H|E\ran=E|E\ran.
\ee
We then can define the time shift operator as
\be
\delta \equiv i \frac{\partial}{\partial E}
\ee
in the energy representation.  Requiring that $\delta$ is hermitian on this Hilbert space then tells us that we must restrict to wave functions $\psi(E)$ which vanish at $E=0$.  

The reader may (rightly) be uncomfortable with this however: there is an old argument due to Pauli that there can be no self-adjoint ``time operator'' which is canonically conjugate to the Hamiltonian in a quantum mechanical system whose energy is bounded from below.  The argument is trivial: if $\delta$ were self-adjoint, then we could exponentiate it to obtain the set of operators $e^{ia\delta}$, which we could use to lower the energy as much as we like, contradicting the lower bound on the energy (see \cite{srinivas1981time} for a more rigorous version of this argument). Therefore our $\delta$, though hermitian, must not be self-adjoint.  In fact this problem is visible already in the classical system: the vector flow on phase space generated by $\delta$ hits the boundary at $H=0$ in finite time.  

These subtleties may be avoided if we instead use the renormalized geodesic distance operator $L$.  Classically this generates a good flow on phase space, so it should correspond to a self-adjoint operator and thus have a basis of (delta-function normalized) eigenstates.  As usual in single-particle quantum mechanics, we can construct the Hilbert space out of $L_2$-normalizable functions of $L$.  The energy eigenstates have wave functions which can be determined from the Schrodinger equation:
\be
-\frac{1}{2\phi_b}\Psi^{\prime\prime}_E(L)+\frac{2}{\phi_b}e^{-L} \Psi_E(L)=E\Psi_E(L).
\ee
The normalizable solutions of this equation with $E>0$ are constructed using modified Bessel functions, in the usual scattering normalization we have
\be\label{scattering}
\Psi_E(L)=\frac{2^{1-2i \sqrt{2E\phi_b}}}{\Gamma(-2i \sqrt{2 E \phi_b})}K_{2i\sqrt{2E\phi_b}}\left(4 e^{-L/2}\right).
\ee
These wave functions decay doubly exponentially at large negative $L$, while at large positive $L$ we have
\be
\Psi_E(L)\approx e^{-i \sqrt{2E\phi_b}L}+R e^{i\sqrt{2E \phi_b}L},
\ee
with reflection coefficient
\be
R=2^{-4i\sqrt{2E\phi_b}}\frac{\Gamma(2i\sqrt{2E\phi_b})}{\Gamma(-2i\sqrt{2E\phi_b})}.
\ee
The Gamma function identity $\Gamma(z)^*=\Gamma(z^*)$ tells us that this reflection coefficient is a pure phase, as is necessary since there is no transmission.  These expressions can be thought of as providing an exact solution of quantum JT gravity with two asymptotic boundaries. 

\subsection{Euclidean Path Integral}
It may seem that given the scattering wave functions \eqref{scattering}, no more needs to be said about quantum JT gravity.  To compare to what we do in higher dimensions however, it is useful to consider how standard Euclidean gravity methods are related to our exact solution.  We begin this discussion by reviewing the Euclidean path integral for JT gravity with \textit{one} asymptotic boundary, on which the time coordinate $t_E$ has periodicity $\beta$.  Namely we sum over geometries with the topology of the disk, and with induced metric 
\be
\gamma_{t_Et_E}=\frac{1}{r_c^2}
\ee
at the boundary.  We then again take the dilaton to obey
\be
\Phi|_{\partial M}=\phi_b r_c,
\ee
and take $r_c\to\infty$ to get asymptotically-AdS boundary conditions.  The Euclidean action is
\be\label{eucS}
-S_E=\int_Md^2x \sqrt{g}\left[\Phi_0 R+\Phi(R+2)\right]+2\int_{\partial M} dx\sqrt{\gamma}\left[\Phi_0K+\Phi(K-1)\right].
\ee
The saddle point for this path integral is the Euclidean Schwarzschild solution
\begin{align}\nonumber
ds^2&=(r^2-r_s^2)dt_E^2+\frac{dr^2}{r^2-r_s^2}\\
\Phi&=\phi_b r,\label{eucschsol}
\end{align}
where smoothness at $r=r_s$ requires
\be
r_s=\frac{2\pi}{\beta}.
\ee
The extrinsic curvature at the boundary is
\be\label{Keq}
K=\frac{r_c}{\sqrt{r_c^2-r_s^2}}=1+\frac{1}{2}\frac{r_s^2}{r_c^2}+\ldots,
\ee
so evaluating the Euclidean action on this solution one finds \cite{Maldacena:2016upp}
\be\label{partZ}
Z[\beta]\equiv \int \mathcal{D}g\mathcal{D}\Phi \,e^{-S_E}\approx e^{4\pi \Phi_0+\frac{4\pi^2 \phi_b}{\beta}}.
\ee
If we interpret this as a thermal partition function, then we can use standard thermodynamic formulas to find the energy and entropy:
\begin{align}\nonumber
\lan H_L\ran&=-\partial_\beta Z=\frac{4\pi^2 \phi_b}{\beta^2}=\phi_b r_s^2=\frac{\Phi_h^2}{\phi_b}\\
S&=\beta\lan H_R\ran+\log Z=4\pi\left(\Phi_0+\Phi_h\right).\label{Seq}
\end{align}
We discuss in section \ref{factsec} below to what extent these can actually be interpreted as thermal energy and entropy, but we note now that one can rewrite the semiclassical result in the suggestive manner
\be\label{Lrho}
Z[\beta]\approx\int_0^\infty dE_L \,\rho_L(E_L)e^{-\beta E_L},
\ee  
with 
\be\label{rhoL}
\rho_L(E_L)\equiv S'(E_L)e^{S(E_L)}\approx e^{4\pi(\Phi_0+\sqrt{\phi_b E_L})}.
\ee


\subsection{Hartle-Hawking State}
Whether or not the Euclidean path integral defines a thermal partition function, we can always use it to define a natural family of states in the Hilbert space of the two-boundary system which we constructed in section \ref{hilbsec} above: these states are labelled by a real parameter $\beta$, and are collectively called the Hartle-Hawking state \cite{Hartle:1976tp,Israel:1976ur}.  They can be interpreted as describing a wormhole connecting the two asymptotic boundaries, where from either side an observer sees a black hole in equilibrium at inverse temperature $\beta$.  


\bfig
\includegraphics[height=4cm]{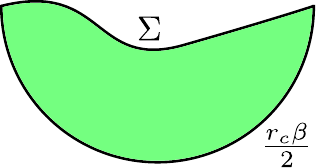}
\caption{The Hartle-Hawking state: we sum over geometries and dilaton configurations with an AdS boundary of length $r_c\beta/2$ and a ``bulk'' boundary $\Sigma$, which we interpret as a time-slice of the two-boundary system. The boundary conditions on $\Sigma$ depend on which basis we wish to compute the wave function in.}\label{hhstatefig}
\efig
The basic idea of the Hartle-Hawking state is illustrated in figure \ref{hhstatefig}.  We can compute the wave function of the Hartle-Hawking state in various bases, the traditional choice is to fix the induced metric on the bulk slice $\Sigma$, together with any matter fields, which computes the wave function in the Wheeler-de Witt representation.  In this section, we will semiclassically compute the wave function of the Hartle-Hawking state in the two bases of the two-boundary Hilbert space, labelled by $E$ and $L$, which we discussed in subsection \ref{hilbsec}.

The $L$ basis calculation is conceptually simpler but technically harder, so we begin with the $E$ basis calculation.    From \eqref{Htot} we know that the energy is a simple function of the value of the dilaton at the bifurcate horizon, $\Phi_h$.  So we need to pick boundary conditions on the bulk slice $\Sigma$ which ensure \textit{(i)} that it passes through the bifurcate horizon and \textit{(ii)} that the dilaton is equal to $\Phi_h$ there.  To achieve these we will require that
\begin{align}\nonumber
n^\mu \partial_\mu \Phi|_\Sigma&=0\\
\Phi|_{\Sigma}&=\Phi_h \sqrt{1+x^2},
\end{align}
where $n^\mu$ is the normal vector to $\Sigma$.  In the second equation we have chosen ``global'' coordinates on the slice, which is not really necessary, but it is perhaps useful to be concrete.  These boundary conditions are consistent with the action variation \eqref{var}, since both boundary terms vanish (remember that the $\Phi\gamma^{\alpha\beta}\delta \gamma_{\alpha\beta}$ and $-\delta\Phi$ terms are not present since this is not an AdS boundary).  More concisely, we want to integrate over geometries with a piece of AdS boundary of length $r_c\beta/2$ and a piece of bulk boundary  $\Sigma$ with vanishing normal derivative of $\Phi$ and $\Phi=\Phi_h$ at its minimum on $\Sigma$.  In the end we may then substitute $\Phi_h=\sqrt{\phi_b E/2}$ to get the wave function in terms of $E$.  We emphasize that here $\Phi_h$ and $\beta$ are \textit{not} related via 
\be\label{peak1}
\Phi_h=2\pi \phi_b/\beta,
\ee 
$\beta$ labels which Hartle-Hawking state we are considering and $\Phi_h$ is the argument in its wave function.  We expect however that \eqref{peak1} should hold at the peak of the wave function.

\bfig
\includegraphics[height=4cm]{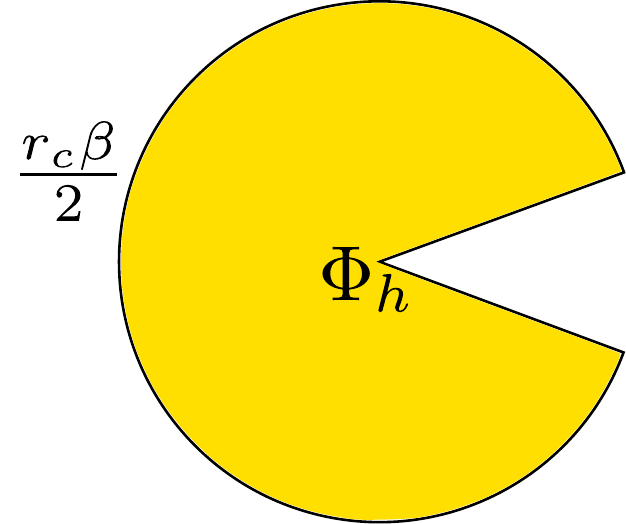}
\caption{Saddle point for the $E$-basis Hartle-Hawking wave function.}\label{pacmanfig}
\efig
The saddle point for this calculation is shown in figure \ref{pacmanfig}, it is a ``sliver'' of the Euclidean Schwarzschild solution \eqref{eucschsol} with $t_E\in(0,\beta/2)$ and $r_s=\frac{\Phi_h}{\phi_b}$.  The kink in the boundary at $r=r_s$ does not violate the boundary conditions since it happens at the minimum of $\Phi$: the derivative of $\Phi$ vanishes in \textit{any} direction there.  To proceed further we need to evaluate the Euclidean action of this solution, but this is complicated by the fact that the solution has corners, which require additional terms in the action not present in \eqref{eucS}.  

We begin our discussion of corner terms by recalling the Gauss-Bonnet theorem in the presence of corners:
\be\label{gbth}
\chi\equiv 2-2g-b=\frac{1}{4\pi}\left(\int_M d^2x \sqrt{g}R+2\int_{\partial M}dx\sqrt{\gamma}K+2\sum_i\left(\pi-\theta_i\right)\right).
\ee
Here $\chi$ is the Euler character, $g$ is the genus, $b$ is the number of boundaries, and $\theta_i$ are the interior opening angles of any corners ($\theta_i=\pi$ means no corner).  These corner contributions can be derived by smoothing out the corner and then taking a limit where the extrinsic curvature $K$ picks up a $\delta$-function contribution.   This suggests that we should upgrade our Euclidean JT action \eqref{eucS} with corner terms
\be
-S_E\rightarrow-S_E+2\sum_i(\Phi_0+\Phi(x_i))\left(\pi-\theta_i \right),
\ee
and indeed this is the correct prescription for corners in the action for a path integral which is describing an overlap of two states.  In using the path integral to compute a wave function however, there are additional corners (such as those in figure \ref{hhstatefig}) which arise from cutting an ``overlap'' type path integral: for these the corner prescription involves $\pi/2-\theta_i$ instead of $\pi-\theta_i$, since we need the corners to cancel when we glue two states together to compute an overlap (see section 3.2 of \cite{Polchinski:1998rq} for some more discussion of this).   If we denote by $C_1$ the set of corners of the first type and $C_2$ the set of corners of the second type, and also the AdS piece of the boundary by $B$, then the full Euclidean action for computing wave functions is
\begin{align}\nonumber
-S_E=&\int_Md^2x \sqrt{g}\left[\Phi_0 R+\Phi(R+2)\right]+2\int_{B+\Sigma} dx\sqrt{\gamma}(\Phi_0+\Phi)K-2\int_Bdx \sqrt{\gamma}\Phi\\
&+2\sum_{i\in C_1}(\Phi_0+\Phi(x_i))\left(\pi-\theta_i \right)+2\sum_{j\in C_2}(\Phi_0+\Phi(x_j))\left(\frac{\pi}{2}-\theta_j \right). \label{eucS2}
\end{align}

Returning now to our saddles for the energy-basis wave function, the saddle in figure \ref{pacmanfig} has corners of both types, but fortunately the corners of type $C_2$ both have $\theta=\frac{\pi}{2}$ so they don't contribute.    The kink at the horizon is a corner of type $C_1$ since it would not contribute if its internal angle $\theta$ were $\pi$, but in fact $\theta$ is 
\be
\theta=\frac{\beta \Phi_h}{2\phi_b}
\ee
so we do have a contribution.  Away from this kink the bulk slice $\Sigma$ is a geodesic, so $K=0$ there.  We thus have
\be
-S_E=\Phi_0\int_M d^2x \sqrt{g}R+2\int_{B} dx\sqrt{\gamma}\left[\Phi_0K+\Phi(K-1)\right]+2(\Phi_0+\Phi_h)\left(\pi-\frac{\beta \Phi_h}{2\phi_b}\right).
\ee
Evaluating this on our saddle point using \eqref{Keq}, and remembering that $t_E$ is integrated from $0$ to $\beta/2$, we find\footnote{The calculation of the $\Phi_0$ terms can be simplified by using the Gauss-Bonnet theorem \eqref{gbth}.} 
\be
-S_E=2\pi(\Phi_0+\Phi_h)-\frac{\beta}{2}\frac{\Phi_h^2}{\phi_b},
\ee
so substituting $\Phi_h=\sqrt{\phi_bE/2}$ as in equation \eqref{Htot} we at last arrive at the semiclassical Hartle-Hawking wave function
\be\label{HHE}
\Psi_{\beta}(E)=\exp\left[2\pi \Phi_0+\sqrt{2\pi^2\phi_b E}-\beta E/4\right].
\ee
As expected, this wave function is peaked when \eqref{peak1} holds.  Moreover if we square it and integrate over $E$, we recover the ``partition function'' $Z[\beta]$ from \eqref{partZ}; in fact we recover $Z[\beta]$ precisely in the representation \eqref{Lrho}.  Here however we are interpreting $Z[\beta]$ not as a thermal trace, but instead as the norm of the Hartle-Hawking state with the normalization produced by the Euclidean path integral.  We further discuss the meaning of \eqref{HHE} in section \ref{factsec} below.

We now proceed to the $L$-basis calculation.\footnote{This calculation is something of an aside to the main points of our paper, so casual readers may wish to skim the remainder of this section.}  We now want $\Sigma$ to be a geodesic of renormalized length $L$, so we now define $\Sigma$ by requiring
\begin{align}\nonumber
K|_\Sigma&=0\\
\gamma_\Sigma&=ds^2,
\end{align}
with the range of $s$ being equal to $L+2\log(2\phi_b r_c)$.  These boundary conditions are again consistent with the variation \eqref{var}, since now $K=0$ and $\delta\gamma_{\mu\nu}=0$ (remember again that the $\Phi\gamma^{\alpha\beta}\delta \gamma_{\alpha\beta}$ and $-\delta\Phi$ terms are not present since $\Sigma$ is not an AdS boundary).  

\bfig
\includegraphics[height=4cm]{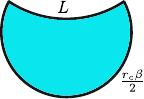}
\caption{Saddle point for the $L$-basis Hartle-Hawking wave function.}\label{HHLfig}
\efig
The saddle points for this calculation are a bit more involved, we want a piece of the Euclidean Schwarzschild geometry \eqref{eucschsol} whose boundary has a piece which is asymptotically AdS, with length $r_c \beta/2$, and a piece which is a geodesic through the bulk, of renormalized length $L$.  We illustrate this in figure \ref{HHLfig}.  There is a two-parameter family of geodesics in this geometry, parametrizing by proper length we have
\begin{align}\nonumber
r(\lambda)&=\sqrt{J^2+r_s^2}\cosh \lambda\\
t_E(\lambda)&=\frac{1}{r_s}\mathrm{arctan}\left(\frac{r_s}{J}\tanh \lambda\right)+t_{E,0},
\end{align}
where $J$ tells us how close our geodesic approaches the center of the disk and $t_{E,0}$ tells us at what value of $t_E$ this closest approach happens.  We can set $t_{E,0}=0$ by convention, so to construct a solution we need to give $J$ and $r_s$ as functions of $L$ and $\beta$ such that our geodesic indeed has renormalized length $L$ and the AdS component of the boundary indeed has length $r_c\beta/2$.  After some algebra, we find that $r_s$ is obtained by solving the equation 
\be\label{aeq}
a=\frac{\sin x}{x},
\ee
with 
\begin{align}\nonumber
x&\equiv \frac{r_s \beta}{4}\\
a&\equiv 4 \phi_b e^{L/2}\beta^{-1},\label{xaeq}
\end{align}
and 
\be
J=\sqrt{e^{-L}\phi_b^{-2}-r_s^2}.
\ee
Note that $a$ is positive, and that we must have $x\in (0,\pi)$.  A unique solution exists provided that $a\leq 1$, or in other words that
\be
\beta\geq4\phi_be^{L/2},
\ee
with $r_s=0$ when this inequality is saturated, and no solution exists for $a>1$.  When $a=2/\pi$, or in other words
\be\label{saddle}
\beta=2\pi \phi_be^{L/2},
\ee
we find that $r_s=\frac{2\pi}{\beta}$, which corresponds to cutting the Euclidean solution \eqref{eucschsol} in two.  This should be what we find is the peak of the wave function.  As $L\to -\infty$ we have $r_s\to \frac{4\pi}{\beta}$.  If we fix $\beta$ and decrease $L$, $r_s$ increases monotonically.

Finally to evaluate the action, we again use \eqref{eucS2}.  There are now no corners of type $C_1$, but we will see that the two corners of type $C_2$ now make a nontrivial contribution.  This is not obvious, since we expect that as $r_c\to\infty$ we have $\theta\to\pi/2$ at each corner for any $\beta$ and $L$, but since $\Phi(x_j)=\phi_b r_c$ we are potentially sensitive to a subleading term in $\theta$ which is $O(1/r_c)$.  Indeed a short calculation tells us that we have
\be
\theta=\frac{\pi}{2}-\frac{r_s}{r_c\tan \left(\frac{\beta r_s}{4}\right)}+\ldots,  
\ee
so there will be a nontrivial corner contribution.  The rest of the action is easy to evaluate, we again have $K=0$ on $\Sigma$ and $R=-2$ in the bulk, so we need only compute the corner terms, the $\Phi_0$ terms, and the terms at the AdS boundary.  The result is
\be
\psi_\beta(L)=\exp\left[2\pi \Phi_0+\frac{8\phi_b}{\beta}\left(x^2+\frac{2x}{\tan x}\right)\right],
\ee
with $x$ determined as a function of $L$ and $\beta$ by solving \eqref{aeq} and using \eqref{xaeq}.  This wave function has a unique maximum at $x=\pi/2$, which from \eqref{xaeq} happens when $r_s=\frac{2\pi}{\beta}$, as expected.  This peak will dominate the integrated square of the wave function, which again is consistent with the saddle point evaluation \eqref{partZ} of $Z[\beta]$.  Near this peak we have
\begin{align}
-S&=\mathrm{constant}-\frac{8\phi_b}{\beta}(x-\pi/2)^2+\ldots\\
&=\mathrm{constant}-\frac{\pi^2\phi_b}{2\beta}\left(L-L_{peak}\right)^2+\ldots,
\end{align}
so the width in $L$ is 
\be
\delta L=\sqrt{\frac{\beta}{\pi^2 \phi_b}},
\ee
which is consistent with the idea that large $\phi_b$ is the semiclassical limit.

Thus we see that the Hartle-Hawking states fit nicely into Hilbert space of the Jackiw-Teitelboim gravity, with (reasonably) simple semiclassical wave functions in the $E$ and $L$ bases.  It would be interesting to extend these calculations to one-loop, in fact the normalization of the Hartle-Hawking state is one-loop exact \cite{Stanford:2017thb}, so the wave function itself might be as well.  

\section{Factorization and the range of the time shift}\label{factsec}
We now return to the interpretation of the single-boundary Euclidean path integral $Z[\beta]$, whose semiclassical value in the Jackiw-Teitelboim gravity is given by \eqref{partZ}.  So far the only Hilbert space interpretation we have given it is as the norm of the unnormalized Hartle-Hawking state in the two-boundary Hilbert space, as produced by the Euclidean path integral without any rescaling.  In AdS/CFT however there is another interpretation for this path integral: following \cite{Maldacena:2001kr}, we can interpret the unnormalized Hartle-Hawking state as corresponding to the unnormalized thermofield double state
\be\label{tfd}
|\psi_{\beta}\ran=\sum_ie^{-\frac{\beta E_i}{2}}|i^*\ran_L|i\ran_R
\ee
in the tensor product Hilbert space of two copies of the boundary CFT.\footnote{Here $|i\ran_R$ are energy eigenstates of the ``right'' CFT and $|i^*\ran_L$ are their conjugates under a two-sided version of CPT which exchanges the two sides, see \cite{Harlow:2014yka} for more explanation of this.}  The one-sided path integral is then the norm of this state,
\be
Z[\beta]=\sum_i e^{-\beta E_i}
\ee
but this is nothing but the one-sided thermal partition function.  Is this interpretation valid in the Jackiw-Teitelboim gravity?

The answer to this last question is no.  The reason is that the Hilbert space of two-boundary Jackiw-Teitelboim gravity, which is just a single-particle quantum mechanics, does not tensor-factorize into a product of one-boundary Hilbert spaces.  Although the Hartle-Hawking state exists, there is no analogue of equation \eqref{tfd}.  Instead we have equation \eqref{HHE}, which we can write in a manner more similar to \eqref{tfd} by labeling states by the one-sided energy $E_L$, which by \eqref{Hs} is half that of the two-sided energy used in \eqref{HHE}, to get
\be
|\psi_\beta\ran\propto \int_0^\infty dE_L e^{2\pi (\Phi_0+\sqrt{\phi_bE_L})-\beta E_L/2}|E_L\ran.
\ee
This is not a state in a tensor-product Hilbert space: indeed there are no states at all where $E_L\neq E_R$, since there is no matter in the pure JT theory all energy is sourced by the bifurcate horizon.  We therefore conclude that \textit{there can be no boundary theory dual to pure quantum Jackiw-Teitelboim gravity}. Were one to have existed, there would have been such a factorization.\footnote{Readers who have casually followed the recent SYK developments may be puzzled, since naively one might have gotten the impression that the two-boundary JT theory should be dual to ``two copies of the Schwarzian theory''.  This is wrong, basically because a single Schwarzian theory in Lorentzian signature does not make sense.  We give the precise statements in the following section.}

There is another interesting illustration of the non-factorization of the JT gravity Hilbert space.  In any tensor product Hilbert space
\be
\mathcal{H}=\mathcal{H}_L\otimes \mathcal{H}_R
\ee
for which the Hamiltonian is a sum of the form
\be
H=H_L\otimes I_R+I_L\otimes H_R,
\ee
we have the partition function identity
\be\label{Z2eq}
Z_{tot}[\beta]\equiv \Tr e^{-\beta H}=\left(\Tr_L e^{-\beta H_L}\right)\left(\Tr_R e^{-\beta H_R}\right)=Z_L[\beta]Z_R[\beta].
\ee
Both sides of this identity are computable in JT gravity, so we can test if it is true.  Assuming factorization, $Z_L$ and $Z_R$ would both be given by the function $Z[\beta]$ we computed in \eqref{partZ}.  $Z_{tot}[\beta]$ we can then attempt to compute by computing the thermal trace in our two-sided Hilbert space.  There is however an immediate problem: since the spectrum of $H$ is continuous, the trace in $Z_{tot}$ is not well-defined.  Let's illustrate this in a simpler example: the quantum mechanics of a free non-relativistic particle of mass $m$ moving on a circle of radius $R$. Momenta is quantized as 
\be
p=\frac{n}{R},
\ee 
so we have a density of states
\be\label{rhoex}
\rho(E)=R\sqrt{\frac{2m}{E}}.
\ee
The thermal partition function is therefore
\be
Z[\beta]=\int_0^\infty dE \,\rho(E)e^{-\beta E}=R\sqrt{\frac{2\pi m}{\beta}}.
\ee
The key point is that the density of states, and therefore the partition function, are divergent in the limit that $R\to\infty$.  In our Jackiw-Teitelboim quantum mechanics with Hamiltonian \eqref{LPH}, the dynamical coordinate $L$ is similarly noncompact, leading to a continuous spectrum with a divergent density of states, so the left hand side of our putative equation \eqref{Z2eq} is divergent while the right hand side is finite.  This is another illustration of the non-factorization of the two-boundary Jackiw-Teitelboim Hilbert space.  

It may seem that we deserved the nonsense we got in attempting to test equation \eqref{Z2eq} in JT gravity, since after all that equation was derived assuming factorization and we already know that the JT Hilbert space doesn't factorize.  But in fact we can use this equation to do something more interesting: we can ask how the theory would need to be modified such that \eqref{Z2eq} would indeed hold.  In other words, what would the two-sided density of states $\rho_{tot}(E)$ need to be such that we indeed had
\be
\int_0^\infty dE \rho_{tot}(E) e^{-\beta E}=Z[\beta]^2?
\ee
In the semiclassical approximation, this can happen only if 
\be
\rho_{tot}(E)\approx \rho_L(E/2)^2\approx e^{2S(E_L)},
\ee
where $\rho_L$ was defined in \eqref{rhoL} and $S$ is the entropy \eqref{Seq}.  In light of \eqref{rhoex}, this equation has a natural interpretation: in a factorized theory obeying \eqref{Z2eq}, our renormalized geodesic observable $L$ cannot really be larger than some length of order $e^{2S(E_L)}$.  From \eqref{Leq} we then also learn that we cannot evolve the Hartle-Hawking state with our total Hamiltonian $H=H_L+H_R$ for times which are longer than of order $e^{2S(E_L)}$ without the Jackiw-Teitelboim description breaking down.  At least then, if not sooner, there must be ``new physics'' in any theory which factorizes.\footnote{The timescale $e^{2S(E_L)}$ is quite natural from the point of view of the proposal that exponentially complex operations should disrupt the structure of spacetime \cite{Harlow:2013tf} \cite{Susskind:2014rva,Brown:2015bva,Susskind:2015toa}: it is the time it takes for the time evolution operator $e^{-iH_L t}$ to reach maximal circuit complexity, and is also the time it takes the thermofield double state to reach maximal state complexity.}   The idea that exponentially long time evolution is in tension with the semiclassical description of the Hartle-Hawking state was also discussed in \cite{Papadodimas:2015xma}.  

In this last argument it may seem that we have gotten ``something for nothing'', since we learned what the range of the time-shift operator $\delta$ must be in a factorized theory using only the Jackiw-Teitelboim path integral.  This is indeed miraculous, but in fact it is the same old miracle by which the Euclidean path integral evaluation of $Z[\beta]$ is able to correctly count black hole microstates using the only the low energy bulk effective action.  This is possible only because that Euclidean path integral is \textit{not} a trace of the Hilbert space of the bulk effective theory with one asymptotic boundary: in fact in JT gravity no such Hilbert space exists.  Given only the bulk theory, the only Hilbert space interpretation we can give to $Z[\beta]$ is as the norm of the unnormalized Hartle-Hawking state: what we have learned here is that factorization is the key assumption which allows us to re-interpret this as a thermal partition function.

\section{Embedding in SYK}
How might we attain a factorizable version of the Jackiw-Teitelboim gravity, in which $Z[\beta]$ is indeed a partition function?  For the $1+1$ Maxwell theory discussed in the introduction, the answer is simple: we need to introduce new matter fields which possess the fundamental unit of $U(1)$ gauge charge.  This modifies Gauss's law such that the electric flux on the left boundary is no longer required to be equivalent to the electric flux on the right boundary, and the Wilson line which stretches from one boundary to the other can be split by a pair of these dynamical charges \cite{Harlow:2015lma}.\footnote{Strictly speaking we also need to introduce a UV cutoff as well since no continuum quantum field theory on a connected space has a factorized Hilbert space. In quantum field theory the question of factorizability is best understood in terms of whether or not the theory obeys the ``split property'', see \cite{hotop} for more discussion of this.}  For gravity we might therefore expect that achieving factorizability is as simple as introducing matter fields, and in some sense this is true. 

In more than two spacetime dimensions, where the global AdS vacuum exists, adding matter would enable us to form one-sided black holes from the collapse of this matter. Therefore the bulk Lagrangian will no longer be no longer UV-complete: the full bulk theory will need to be able to count the microstates of those black holes in Lorentzian signature.  In two dimensions, the space must end somewhere, however we may add collapsing matter on top of a smaller two-sided black hole to produce a larger one. Then it will again be the case that the full bulk theory must be able to account for the exponentially large number of additional microstates. 
What we really need then is to find a holographic boundary description, where we understand the theory as a large-$N$ quantum mechanics living on the disconnected space $\mathbb{R}\sqcup \mathbb{R}$.

  So far the only known explicit examples of this are based on the Sachdev-Ye-Kitaev model \cite{Sachdev:1992fk,Kitaev,Polchinski:2016xgd,Maldacena:2016upp,Maldacena:2016hyu,Kitaev:2017awl}.  These examples unfortunately have a large number of light matter fields, which cause bulk locality to break down at the AdS scale, but they do also have a Jackiw-Teitelboim sector which decouples from all that at low temperatures.\footnote{Having locality break down at the AdS scale seems to be a common feature of all ``exactly solvable'' models of AdS/CFT, see \cite{Klebanov:2002ja,Pastawski:2015qua} for other examples.  The main advantage of SYK is that there is a parametric limit, low temperature, where the gravitational sector decouples and the theory stays solvable.}  In Euclidean signature this was shown in \cite{Maldacena:2016hyu,Jensen:2016pah,Engelsoy:2016xyb,Maldacena:2016upp,Kitaev:2017awl}, in this section we sketch the (fairly trivial) modifications which are needed to give the analogous argument in Lorentzian signature.  
	
The SYK model is a collection of $N$ Majorana fermions $\chi^a$, interacting with Hamiltonian 
\be
H=-\frac{1}{4!}\sum_{a,b,c,d}J_{abcd}\chi^a\chi^b\chi^c\chi^d,
\ee
where the antisymmetric tensor $J_{abcd}$ represents disorder drawn at random from the Gaussian ensemble
\be
P[J]\propto \exp\left[-\frac{N^3}{12J^2}\sum_{a<b<c<d}(J_{abcd})^2\right].
\ee
The Lagrangian of the SYK model is
\be
L=\frac{i}{2}\sum_a\chi^a \dot{\chi}^a+\frac{1}{4!}\sum_{a,b,c,d}J_{abcd}\chi^a\chi^b\chi^c\chi^d.
\ee
We are interested in the Lorentzian path integral for two copies of this model, so our dynamical variables will be $2N$ Majorana fermions $\chi_i^a$, where $a$ runs as before from $1$ to $N$, while $i$ is equal to either $L$ or $R$ and tells us which copy we are talking about.  We will take the disorder $J_{abcd}$ to be the same for each copy, since the ``real'' model corresponds to a single instantiation of the disorder and we want the same Hamiltonian on both sides.  

The large-$N$ solution of this model begins with an assumption, justified by numerics, that we can view the disorder as ``annealed'' rather than ``quenched''. This means that we can integrate over it directly in the path integral rather than waiting until we compute observables to average over it.  We are then interested in evaluating the Lorentzian path integral
\begin{align}\nonumber
Z=\int \mathcal{D}J \mathcal{D}\chi \exp \Bigg[&-\frac{1}{2}\int dt\sum_{a,i} \chi^a_i \dot{\chi}^a_i +i \sum_{a<b<c<d}J_{abcd}\int dt \sum_i\left(\chi_i^a\chi_i^b\chi_i^c\chi_i^d\right)\\
&-\frac{N^3}{12J^2}\sum_{a<b<c<d}(J_{abcd})^2\Bigg].
\end{align}
At large $N$ this integral can be done by a version of the Hubbard-Stratonivich transformation \cite{stratonovich1957method,hubbard1959calculation}.  We first integrate out the disorder, to arrive at
\be
Z=\int \mathcal{D}\chi \exp\left[-\frac{1}{2}\int dt\sum_{a,i} \chi^a_i \dot{\chi}^a_i-\frac{3J^2}{N^3}\sum_{a<b<c<d}\left(\sum_i\int dt \chi_i^a\chi_i^b\chi_i^c\chi_i^d\right)^2\right].
\ee
We then ``integrate in'' bilocal auxilliary fields $\Sigma_{ij}(t,t')$ and $G_{ij}(t,t')$ such that 
\begin{align}\nonumber
Z=\int \mathcal{D}G\mathcal{D}\Sigma \mathcal{D}\chi \exp\Bigg[&-\frac{1}{2}\int dt\sum_{a,i} \chi^a_i \dot{\chi}^a_i\\\nonumber
&+\frac{iN}{2}\sum_{i,j}\int dt \int dt' \Sigma_{ij}(t,t')\left(G_{ij}-\frac{1}{N}\sum_a \chi^a_i(t)\chi_j^a(t')\right)\\
&-\frac{J^2 N}{8}\sum_{i,j}\int dt \int dt'G_{ij}(t,t')^4\Bigg].
\end{align}
Finally we can then integrate out the fermions, which are now Gaussian, arriving at\footnote{We have suppressed details related to renormalization and the $i\epsilon$ prescription here, see \cite{Kitaev:2017awl} for a discussion of the former in the Euclidean case.}
\be
Z=\int \mathcal{D}G\mathcal{D}\Sigma\exp\left[iS(G,\Sigma)\right],
\ee
with the bilocal effective action $S(G,\Sigma)$ given by
\be
S(G,\Sigma)=-\frac{iN}{2}\log \det \left(\delta_{ij}\partial_{t'}-i\Sigma_{ij}\right)+\frac{N}{2}\sum_{i,j}\int dt \int dt'\left(\Sigma_{ij}G_{ij}+\frac{iJ^2}{4}G_{ij}^4\right).
\ee
Here the determinant is defined for matrices with both $ij$ and $tt'$ indices.  The equations of motion are
\begin{align}\nonumber
\Sigma_{ij}&=i J^2 G_{ij}^3\\
\partial_{t'}-i\Sigma&=G^{-1},\label{sykeom}
\end{align}
where we have used matrix notation in the second equation.  

Now the key observation is that at low energies compared to $J$, we can ignore the time derivative in equation \eqref{sykeom}, in which case these equations of motion become invariant under the reparametrization symmetry $\mathrm{diff}(\mathbb{R})\times \mathrm{diff}(\mathbb{R})$:
\begin{align}\nonumber
\Sigma_{ij}'(t_1',t_2')&=(f'_i(t_1)f'_j(t_2))^{-3/4}\Sigma_{ij}(t_1,t_2)\\
G'_{ij}(t_1',t_2')&=(f'_i(t_1)f'_j(t_2))^{-1/4}G_{ij}(t_1,t_2).\label{repam}
\end{align}
In these transformations the primed and unprimed times are related by
\begin{align}\nonumber
t_1'&=f_i(t_1)\\
t_2'&=f_j(t_2).
\end{align}
Most of this symmetry however will be spontaneously broken by any particular saddle point solution $G_{ij}^c$, $\Sigma_{ij}^c$. In Lorentzian signature we are interested in excitations about the zero-temperature thermofield double state: this state will be nontrivial since at large $N$ the SYK model has a large vacuum degeneracy. The equations of motion \eqref{sykeom} can be solved without too much difficulty by going to momentum space, the result is that the matrix $G_{ij}(t,t')$ is nothing but the boundary two-point function of a Majorana fermion in global $AdS_2$, with metric \eqref{globalsol}.  The $ij$ indices tell us which boundary each of the two fermions is on.  This two-point function is invariant under only under the $SL(2,\mathbb{R})$ subgroup of $\mathrm{diff}(\mathbb{R})\times \mathrm{diff}(\mathbb{R})$ which is inherited from Lorentz transformations in the embedding space \eqref{embedmet} (since we have fermions the Lorentz group is $SL(2,\mathbb{R})$ instead of $SO(1,2)$.  Thus at low energies we expect a set of zero modes taking values in 
\be
\Big(\mathrm{diff}(\mathbb{R})\times \mathrm{diff}(\mathbb{R})\Big)/SL(2,\mathbb{R}).
\ee 
These zero modes, let's call them $\phi_n$, will be lifted by finite $J$ effects, so they will have an effective action of the form
\be
S(\phi_n)=\frac{N}{J} s(\phi_n):
\ee 
the lowest-order in derivatives action with this symmetry is two copies of the Schwarzian action,
\be
s(\phi_n)\propto \sum_i\int dt\left\{f_i(t),t\right\},
\ee
where $f_i$ are our two diffeomorphisms of $\mathbb{R}$ and then we quotient by $SL(2,\mathbb{R})$ to get the action for the $\phi_n$.  The classical solutions of the equations of motion obtained by varying this action are a pair diffeomorphisms $f_i(t)$ induced by distinct boundary $SL(2,\mathbb{R})$ transformations, identified modulo the joint $SL(2\mathbb{R})$ induced by isometries of global $AdS_2$.  We then can simply note that in \cite{Maldacena:2016upp} precisely this theory, two copies of the Schwarzian theory with a mixed $SL(2,\mathbb{R})$ gauged, was derived from the Lorentzian JT theory with two asymptotic boundaries (see also \cite{Jensen:2016pah,Engelsoy:2016xyb}.\footnote{The basic idea is that if we solve the metric equation but not yet the $\Phi$ equation, then the functions $f_i(t)$ keep track of where the two boundaries where $\Phi=\phi_b r_c$ are located in $AdS_2$.  We should quotient by the embedding space isometry group $SL(2,\mathbb{R})$, which acts nontrivially on both $f_L(t)$ and $f_R(t)$.  The Schwarzian actions arise from the boundary terms in the action \eqref{JTact}.}  This completes the derivation of the Lorentzian JT theory from the SYK model.  
%

There are two important observations about this derivation:
\bi
\item[(1)] We see that JT gravity is \textit{not} equal to ``two copies of the Schwarzian theory'', at least not in the naive sense of having a tensor product of two sensible Lorentzian theories.  There is a tensor product in a larger unphysical Hilbert space obtained by quantizing pairs of diffeomorphisms, but we must quotient by the subgroup $SL(2,\mathbb{R})$ which mixes the two so the physical Hilbert space does not factorize.  Doing this quotient separately for each diffeomorphism would have led to an empty theory.  
\item[(2)] This theory \textit{is} embedded into a larger Hilbert space which does tensor-factorize, that of two copies of the SYK model (with a fixed instantiation of the disorder).  In describing the low energy sector however, we found ourself needing to use ``left-right'' degrees of freedom which, in the original SYK variables, have the form
\be\label{GLR}
G_{LR}(t,t')=\frac{1}{N}\sum_a\chi_L^a(t)\chi_R^a(t').
\ee
\ei
This last equation is quite interesting from the point of view \cite{Harlow:2015lma}: it is a gravitational version of the procedure of splitting a Wilson line with a pair of dynamical charges.  Note in particular that although the bulk fermions created by $\chi^a_i$ are not present in the JT gravity, we still need them to express the JT degrees of freedom within the SYK description.  This was one of the main lessons of \cite{Harlow:2015lma}: in the presence of bulk gauge fields, mapping low-energy bulk operators into the boundary theory can require heavy bulk degrees of freedom which do not otherwise appear in the low-energy effective action.\footnote{The form of \eqref{GLR} is quite similar to the equation 5.14 in \cite{Harlow:2015lma} for the emergent Wilson link in the $\mathbb{CP}^{N-1}$ nonlinear $\sigma$ model.  In both cases we have a large-$N$ average over bilinears of microscopic charges.  That model thus seems to be quite a good model of emergent gravity in this particular case.}  
 
\section{Conclusion}
One important lesson of Jackiw-Teitelboim gravity is that bulk quantum gravity can make sense with a local Lagrangian.  Indeed we have nonperturbatively constructed the Hilbert space and dynamics of the two-boundary Jackiw-Teitelboim gravity, and we have shown that many calculations are feasible within this simple setting.  There are many more calculations which we did not attempt, two which we expect would be quite interesting are extending our calculation of the Hartle-Hawking wave function to one loop (and perhaps beyond), and repeating our analysis for the supersymmetric version of Jackiw-Teitelboim gravity.  

We believe that the basic reason for why the JT Lagrangian leads to a well-defined bulk theory of quantum gravity is precisely that the Hilbert space it constructs doesn't factorize: even though it has wormhole solutions, it does not have black hole microstates.  We have seen that the usual computations of black hole thermodynamics can all be given ``non-thermodynamic interpretations'' within this theory, with in particular the Euclidean one-boundary path integral being interpreted as the normalization of the unnormalized Hartle-Hawking state rather than a thermal partition function. 

\bfig
\includegraphics[height=5cm]{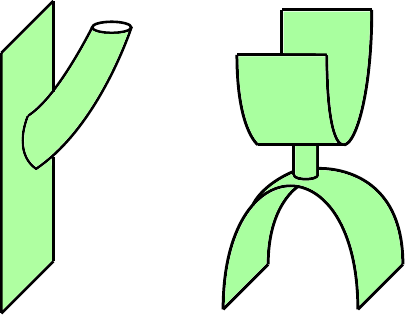}
\caption{Topology-changing Euclidean configurations.  Our quantization of the Jackiw-Teitelboim theory does not include such processes, consistent with what the Lorentzian path integral would predict.  But perhaps it should?}\label{circlesplitfig}
\efig
One important issue which we have not explored in detail is the role of topologically nontrivial configurations in the Euclidean path integral of the Jackiw-Teitelboim theory.  Off-shell field configurations certainly exist where the spacetime evolves from a spatial line interval to a line interval plus any number of circles, and if there are more than two asymptotic boundaries then additional ``rewiring configurations'' are possible, which change which pairs of asymptotic boundaries are connected.  We illustrate two such configurations in figure \ref{circlesplitfig}.  Such topology-changing configurations are not present in the Lorentzian path integral, at least not if we define it to include sums only over globally-hyperbolic (in the AdS sense) geometries, and there are also usually not real Euclidean solutions with these topologies.  Moreover the SYK model does not seem to have a discrete infinity of additional states associated to including an arbitrary number of spatial circles.  Nonetheless it would be good to understand in what circumstances we can or should give a physical interpretation to these configurations, for example in AdS/CFT topology-changing Euclidean configurations are sometimes needed to reproduce known CFT results \cite{Jafferis:2017tiu}.   We leave exploration of this question to future work, but we emphasize that until it is addressed we cannot really claim to completely understand the bulk path integral formulation of Jackiw-Teitelboim gravity.  

It is interesting to consider if such a self-contained theory of gravity is possible in higher dimensions: for $3+1$ dimensions and higher we expect that the answer is no, since once there are propagating gravitons these are already enough to make black holes whose microstates must be counted.  But what about $2+1$?  In fact we suspect that pure Einstein gravity in $2+1$ dimensions with negative cosmological constant, and also its supersymmetric extension, give two more examples of nontrivial bulk theories of quantum gravity which make sense as local path integrals but do not have CFT duals.  Here are some features which resemble those of JT gravity:\footnote{Another resemblance is that both theories perturbatively have first-order reformulations as topological gauge theories, perhaps the puzzling features of this reformulation in $2+1$ dimensions could be better understood by studying their $1+1$ dimensional avatars.}
\bi
\item All UV divergences in their path integrals can be absorbed by simple renormalizations of $G$ and $\Lambda$, so they are ``secretly renormalizable'' \cite{Witten:1988hc,Witten:2007kt}.
\item They have two-boundary wormholes, namely the BTZ solution \cite{Banados:1992wn}, and thus have semiclassical Hartle-Hawking states, whose normalization gives the one-boundary Euclidean path integral with boundary $\mathbb{S}^1\times \mathbb{S}^1$.
\item There are no propagating degrees of freedom in the bulk, but the quantum mechanics of the time-shift operator and the Hamiltonian still exist, and thus give a nontrivial dynamics to the two-boundary system.  This is now in addition to the boundary gravitons which are present even with one asymptotic boundary. 
\item The one-boundary theory, while no longer trivial because of boundary gravitons and topologically nontrivial black hole geometries, does not have nearly enough states to account for the Bekenstein-Hawking entropy which the normalization of the Hartle-Hawking state would have predicted \cite{Maloney:2015ina}. 
\ei
Thus we conjecture that a complete quantization of pure Einstein gravity with negative cosmological constant (and its supersymmetric extension) should be possible using bulk path integral methods in $2+1$ dimensions.  The existence of the BTZ ``black hole'' is no obstruction to this, since it should be interpreted as a wormhole instead of a one-sided black hole.  As we found in JT gravity, we expect that the two-boundary Hilbert space will not factorize due to the nonlocal consequences of the diffeomorphism constraints, which therefore would immediately imply that this Hilbert space cannot arise from that of a boundary CFT on a disconnected space.  These conjectures are consistent with the results of Maloney and Witten, who computed the one-boundary partition function exactly and saw that it did not have the form of a thermal trace \cite{Maloney:2007ud}.\footnote{A similar analysis was done for JT gravity in \cite{Stanford:2017thb}, where it was again observed that the result is not consistent with a thermal partition function.}  There has been a fair bit of worry about how to ``fix'' this, for example by including complex saddle points or additional Planckian degrees of freedom, but inspired by JT gravity our proposal is instead that this is simply the right answer!   Pure gravity in $2+1$ dimensions with negative cosmological does exist, but it doesn't have a dual CFT.\footnote{The reader may object that this isn't what they mean by pure gravity, since we should allow for new physics at the Planck scale.  In some sense this is semantic, but if we allow this then how could the question of whether ``pure quantum gravity'' as a unique entity exists be well-defined?  What is the appeal of a renormalizable theory of quantum gravity if we are going to treat it as an effective theory anyways?  The nontrivial question which seems most natural, which was also the one studied in \cite{Maloney:2007ud}, is whether or not the assumed-to-be-well-defined bulk path integral is really equivalent to some boundary CFT without further modification.  \cite{Maloney:2007ud} already gave strong evidence that the answer is no, our only contribution here is to emphasize that this is \textit{not} an argument that the bulk theory is inconsistent on its own terms.}  This proposal clearly needs more scrutiny before it should be accepted, but with the JT theory to guide us, where many of the same issues arise in simpler guise, it seems to be time for another shot.  

\paragraph{Acknowledgments} We thank Adam Brown, Shamit Kachru, Alexei Kitaev, Hong Liu, Juan Maldacena, Hirosi Ooguri, Steve Shenker, Josephine Suh, Leonard Susskind, Douglas Stanford, Andy Strominger, Xiaoliang Qi, and Edward Witten for helpful conversations.   We also thank Henry Lin for pointing out a mistake in an earlier version of this paper.  DH is supported by the Simons Foundation ``It-from-Qubit'' Collaboration and the MIT Department of Physics. The work of DJ is supported in part by NSFCAREER grant PHY-1352084.

\bibliographystyle{jhep}
\bibliography{bibliography}
\end{document}